\documentclass[12pt]{article}
\usepackage[dvips]{graphicx}
\usepackage{amsmath,amssymb,amsfonts}
\renewcommand{\baselinestretch}{1.3}
\begin{document}

\title{Kaon regeneration in perturbation theory}
\author{V.I. Nazaruk\\
Institute for Nuclear Research of RAS, 60th October\\
Anniversary Prospect 7a, 117312 Moscow, Russia.*}

\date{}
\maketitle
\bigskip

\begin{abstract}
$K^0\bar{K}^0$ transitions in a medium followed by decay and regeneration 
of $K^0_{S}$-component are considered by means of perturbation theory. It is 
shown that in the previous calculations the problem different from regeneration 
is solved. 
\end{abstract}

\vspace{5mm}

\vspace{5mm}

\vspace{1cm}

*E-mail: nazaruk@inr.ru

\newpage
\setcounter{equation}{0}
\section{Introduction}
We consider the $K^0\bar{K}^0$ transitions in the medium followed by decay:
\begin{equation}
(K^0-\mbox{medium})\rightarrow (\bar{K}^0-\mbox{medium})\rightarrow (f-\mbox{medium})
\end{equation}
($f$ are the decay products) by means of perturbation theory. The regeneration is
considered as well. 

The effect of kaon regeneration is known since the 1950s. The motive of our paper is as follows.
In the previous calculations [1-5] the amplitude of regenerated 
$K^0_{S}$ is proportional to the factor $f_ {21}/\Delta m$. Here $f_{21}=f-
\bar{f}$, $\Delta m=m_L-m_S$, where $m_L$ and $m_S$ are the masses of stationary 
states, $f$ and $\bar{f}$ are the forward scattering amplitudes of $K^0$ and 
$\bar{K}^0$, respectively. As this takes place, $f_{21}$ and $\Delta m$ are the crucial 
values which define the process speed. In our opinion the inverse dependence takes place.

Indeed, in our calculation the amplitude of regenerated $K^0_{S}$ is 
$\epsilon /V\sim \Delta m/f_{21}$ (see Eqs. (25) and (26)). We show that the reason of 
disagreement is that in the standard calculations [1-5] the non-coupled equations of motion 
are used instead of system of coupled ones (see Eq. (27) and below). This means that the 
process model is wrong.

The regeneration arises from difference between the $K^0N$- and $\bar{K}^0N$-intereactions. 
We propose the simple model which is typical for the theory of the multistep processes.
We specify the initial and final states and interaction Hamiltonian. The calculation is 
performed by means of perturbation theory.

In Sect. 2 the $K^0\bar{K}^0$ transitions in the medium are considered. These 
results are used in Sect. 3, where the main calculation of regeneration is 
performed. Our and previous calculations are compared in Sect. 4. In Sect. 5 we
touch on briefly the approach based on exact solution of equations of motion.  
Section 6 contains the conclusion.

\section{$K^0\bar{K}^0$ transitions in the medium}
The $K^0\bar{K}^0$ transitions in the medium are of independent interest. 
Besides, the results obtained are used in the model of regeneration.

\subsection{Equations of motion} 
To understand the origin of the factor $\Delta m/f_{21}$ we write the
coupled equations for zero momentum $K^0$ and $\bar{K}^0$ in the medium:
\begin{equation}
(i\partial_t-M)K^0=\epsilon \bar{K}^0,
\end{equation}
\begin{equation}
[(i\partial_t-(M+V)]\bar{K}^0=\epsilon K^0,
\end{equation}
where
\begin{eqnarray}
M=m_{K^0}+U_{K^0}-i\Gamma _{K^0}/2,\nonumber\\
V=(m_{\bar{K}^0}-m_{K^0})+(U_{\bar{K}^0}-U_{K^0})-(i\Gamma _{\bar{K}^0}/2-i\Gamma
_{K^0}/2).
\end{eqnarray}
Here $\epsilon =(m_L-m_S)/2$ is a small parameter, $U_{K^0}$ and $U_{\bar{K}^0}$
are the potentials of $K^0$ and ${\bar{K}^0}$, respectively. Below we put 
$m_{K^0}=m_{\bar{K}^0}=m$ and $\Gamma _{K^0}= \Gamma _{\bar{K}^0}=\Gamma $, 
where $m$ and $\Gamma $ are the mass and decay width of $K^0$, respectively. 
 
The initial conditions are $K^0(t=0)=1$ and $\bar{K}^0(t=0)=0$. In the lowest order
in $\epsilon $
\begin{eqnarray}
(i\partial_t-M)K^0(t)=0,\nonumber\\
K^0(t)=\exp (-iMt).
\end{eqnarray}
Substitution of $K^0(t)$ in (3) gives
\begin{eqnarray}
\bar{K}^0(t)=\frac{\epsilon }{V}[\exp (-iVt)-1]K^0(t),\nonumber\\
V=U_{\bar{K}^0}-U_{K^0}=\frac{2\pi }{m}Nf_{21},
\end{eqnarray}
where $N$ is the number of nucleons in a unit of volume.

Since ${\rm Im}U_{K^0}\ll {\rm Im}U_{\bar{K}^0}$, we put ${\rm Im}U_{K^0}=0$
for simplicity. The probability of finding a $\bar{K}^0$ in a time $t$ is 
found to be 
\begin{equation}
I(\bar{K}^0)=\frac{\epsilon ^2}{\mid V\mid ^2}[1-2\cos({\rm Re}Vt)e^{{\rm
Im}Vt}+e^{2{\rm Im}Vt}]e^{-\Gamma t}.  
\end{equation}
$I(\bar{K}^0)$ is damped out due to decay: $I(\bar{K}^0)\sim e^{-\Gamma t}$. The
same is true for $ K^0(t)$: $\mid K^0(t)\mid ^2=e^{-\Gamma t}$.

If $V\rightarrow 0$, $I(\bar{K}^0)$ coincides with the free-space $K^0\bar{K}^0$
transition probability [5]. As seen from (6), the factor $\epsilon /V\sim \Delta
m/f_{21}$ is beyond question.

\begin{figure}[!h]
  {\includegraphics[height=.2\textheight]{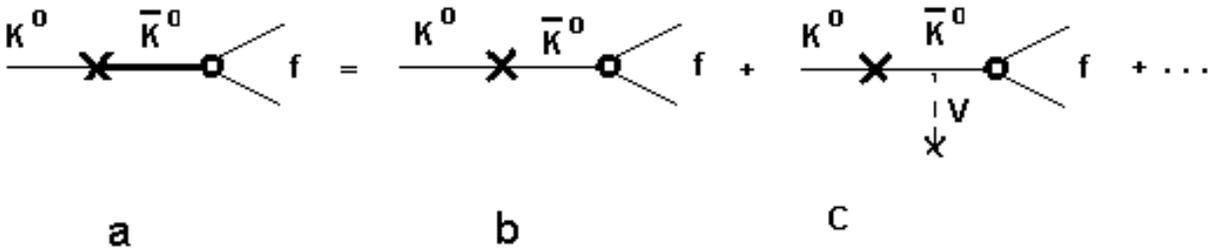}}
  \caption{$K^0\bar{K}^0$ transition in the medium followed by decay.}
\end{figure}

\subsection{Diagram technique}
Let us consider the process (1) (see Fig. 1a). The physical aspects of the problem
are primary goal of this paper. To draw the analogy to the well-studied 
$n\bar{n}$ transitions we consider the non-relativistic problem. The unperturbed 
and interaction Hamiltonians are
\begin{eqnarray}
H_0=-\nabla^2/2m+ U_{K^0},\nonumber\\
H_I=H_{K^0\bar{K}^0}+H_W+V,\nonumber\\
H_{K^0\bar{K}^0}=\int d^3x(\epsilon \bar{\Psi }_{\bar{K}^0}(x)\Psi _{K^0}(x)+H.c.).
\end{eqnarray}
Here $H_{K^0\bar{K}^0}$ and $H_W$ are the Hamiltonians of the $K^0\bar{K}^0$
conversion and decay of the $K$-mesons, respectively; $\bar{\Psi }_{\bar{K}^0}$ and
$\Psi _{K^0}$ are the fields of $\bar{K}^0$ and $K^0$, respectively. The background 
field $U_{K^0}$ is included in the wave function of $K^0$:
\begin{equation}
K^0(x)=\Omega ^{-1/2}\exp (-ipx),
\end{equation}
where $p=(E,{\bf p})$ is the 4-momentum of $K^0$; $E={\bf p}^2/2m+{\rm Re}U_{K^0}$. 
The calculation given below is unaffected by absorption and decay of incident 
particle ($K^0$ in this case) and so these effects  will be taken into account in 
the final result (see last paragraph of Sect. 3).

In the lowest order in $H_{K^0\bar{K}^0}$ the amplitude $M_1$ of the process (1) is 
\begin{eqnarray}
M_1=\epsilon G(\bar{K}^0)M_{d}(\bar{K}^0\rightarrow f),\nonumber\\
G(\bar{K}^0)=\frac{1}{E_1-{\bf p}^2_1/2m-U_{\bar{K}^0}+i0}=-\frac{1}{V}.
\end{eqnarray}
Here $M_{d}(\bar{K}^0\rightarrow f)$ is the in-medium amplitude of the decay
$\bar{K}^0\rightarrow f$, $G(\bar{K}^0)$ is the propagator of the $\bar{K}^0$,
$E_{1}$ and ${\bf p}_{1}$ are the energy and momentum of $\bar{K}^0$;
$E_1=E$, ${\bf p}_1={\bf p}$. The amplitude $M_{d}(\bar{K}^0\rightarrow f)$ 
contains all the orders in $H_W$. 
For the process width $\Gamma _1$ one obtains
\begin{eqnarray}
\Gamma _1=N_1\int d\Phi \mid\!M_1\!\mid ^2=\frac{\epsilon ^2}{\mid V\mid ^2}
\Gamma (\bar{K}^0\rightarrow f),\nonumber\\
\Gamma (\bar{K}^0\rightarrow f)=N_1\int d\Phi \mid\!M_{d}(\bar{K}^0\rightarrow
f)\!\mid ^2,
\end{eqnarray}
where $\Gamma (\bar{K}^0\rightarrow f)$ is the width of the decay
$\bar{K}^0\rightarrow f$. The normalization multiplier $N_1$ is the same for $\Gamma
_1$ and $\Gamma (\bar{K}^0\rightarrow f)$.

The process probability $W_1(t)$ is determined by exponential decay law:
$W_1(t)=1-\exp (-\Gamma _1t)$. If $\Gamma _1t\ll 1$,
\begin{eqnarray}
W_1(t)\approx \Gamma _1t=\frac{\epsilon ^2}{\mid V\mid ^2}
W(\bar{K}^0\rightarrow f),\nonumber\\
W(\bar{K}^0\rightarrow f)=\Gamma (\bar{K}^0\rightarrow f)t.
\end{eqnarray}
Here $W(\bar{K}^0\rightarrow f)$ is the probability of decay $\bar{K}^0\rightarrow f$
in a time $t$.

If $V=0$, $M_1\sim 1/0$ (see Fig. 1b). The same is true for the free-space process.
In this case $U_{K^0}=V=0$ and the process amplitude $M^v_1\sim 1/0$. This problem
should be solved; otherwise the approach under study can be rejected.

For the $n\bar{n}$ transition the similar picture takes place. These are infrared 
singularities conditioned by zero momentum transfer in the $K^0\bar{K}^0$ 
($n\bar{n}$) transition vertex. For solving the problem the field-theoretical 
approach with finite time interval has been proposed [6-8]. It is infrared free.
The problem is formulated on the finite time interval as for the Eqs. (2), (3).

For the total free-space probability $W_t$ of the process (1) the calculation identical 
to that of Ref. [8] gives
\begin{equation}
W_t(t)=\epsilon ^2t^2.
\end{equation}
$W_t$  coincides with the well-known result for the free-space oscillations. 
(The $t$-dependence of (12) and (13) is different. This is well-known fact 
pertaining to exponential decay law.)

An important point is that for the $n\bar{n}$ transition the calculation [9] identical to the calculation of $W_1(t)$
gives the result which coincides with that obtained by means of equations of motion (see 
Eqs. (15) and (16) of Ref. [9] for the case of small $\Gamma $). This fact as well as 
reproducing of the vacuum limit (see (13)) are the verification of the approach under consideration.

We would like to emphasize that $M_1\sim \epsilon /V \sim \Delta m/f_{21}$ as well 
as $\bar{K}^0(t)$.

\begin{figure}[!h]
  {\includegraphics[height=.2\textheight]{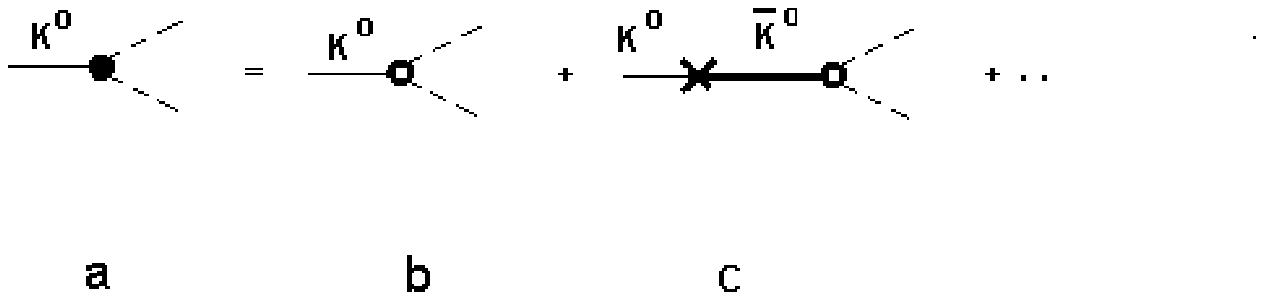}}
\caption{Decay $K^0\rightarrow 2\pi $ in the medium. (b) and (c) correspond to 
zero and first orders in $H_{K^0\bar{K}^0}$, respectively.} 
\end{figure}

\section{Regeneration}
We consider the regeneration of $K^0_{S}$-component. Let $K^0_{L}$ falls on the
plate. For definitess, the decay into $2\pi $ is registered [10].
Since $K^0N$- and $\bar{K}^0N$-intereactions are known, we go into basis 
$(K^0,\bar{K}^0)$ using the vacuum relation 
\begin{equation}
\left.\mid\!K^0_L\right>=\left.(\mid\!K^0\right>+\left.\mid\!\bar{K}^0\right>)/\sqrt{2}.
\end{equation}
Here $\left.\mid\!K^0\right>$, $\left.\mid\!\bar{K}^0\right>$ and $\left.\mid\!K^0_L\right>$ are the 
vacuum states of $K^0$, $\bar{K}^0$ and $K^0_L$, respectively.

To this point kaon is in the vacuum. In (14) the all states are the vacuum states.
We consider $K^0$-component in (14). The spatial wave function of the 
$K^0$ in the vacuum $K^0_v(x)$ is
\begin{eqnarray}
K^0_v(x)=\Omega ^{-1/2}\exp (-iEt+i{\bf p}_v{\bf x}),\nonumber\\
E={\bf p}_v^2/2m.
\end{eqnarray}
For the wave function in the medium $K^0_m(x)$ we have
\begin{eqnarray}
K^0_m(x)=\Omega ^{-1/2}\exp (-iEt+i{\bf p}{\bf x}),\nonumber\\
E={\bf p}^2/2m+{\rm Re}U_{K^0}.
\end{eqnarray}
The sole difference between $K^0_v(x)$ and $K^0_m(x)$ is that $p\ne p_v$.

The coefficient of reflection is small:
\begin{equation}
\left(\frac{p-p_v}{p+p_v}\right)^2\ll 1.
\end{equation}

For the $K^0$ with the in-medium wave function (16) one should describe the process (1) 
with the $\pi \pi $ in the final state. The wave function of initial state 
and interaction Hamiltonian are given by (16) and (8), respectively. 
The perturbation theory in $H_{K^0\bar{K}^0}$ is used.  

For the amplitude of the process $K^0\rightarrow \pi \pi $ in the medium 
$M'(K^0\rightarrow \pi \pi )$ we have
\begin{equation}
M'(K^0\rightarrow \pi \pi )=M_d(K^0\rightarrow \pi \pi)+\epsilon 
G(\bar{K}^0)M_d(\bar{K}^0\rightarrow \pi \pi)
\end{equation}
(see Fig. 2). $M_d(K^0\rightarrow \pi \pi )$ and Fig. 2b correspond to zero 
order in $H_{K^0\bar{K}^0}$. They describe the direct decay $K^0\rightarrow 
\pi \pi $. Figure 2c and term $\epsilon G(\bar{K}^0)M_d(\bar{K}^0\rightarrow 
\pi \pi)$ correspond to the first order in $H_{K^0\bar{K}^0}$ and all the
orders in $V$ and $H_W$.

Similarly, for the $\bar{K}^0$ one obtains
\begin{equation}
M'(\bar{K}^0\rightarrow \pi \pi )=M_d(\bar{K}^0\rightarrow \pi \pi)+
\epsilon G(K^0)M_d(K^0\rightarrow \pi \pi).
\end{equation}
In accordance with (14), the total in-medium amplitude 
$M'(K^0_{L}\rightarrow \pi \pi )$ is
\begin{eqnarray}
M'(K^0_{L}\rightarrow \pi \pi )=\frac{1}{\sqrt{2}}[M'(K^0\rightarrow 
\pi \pi )+ M'(\bar{K}^0\rightarrow \pi \pi )]=\nonumber\\
\frac{1}{\sqrt{2}}[M_d(K^0\rightarrow \pi \pi )+\epsilon G(\bar{K}^0)
M_d(\bar{K}^0\rightarrow \pi \pi)+M_d(\bar{K}^0\rightarrow \pi \pi )+
\epsilon G(K^0)M_d(K^0\rightarrow \pi \pi)],\nonumber\\
G(\bar{K}^0)=-G(K^0)=-1/V.
\end{eqnarray}
Our interest is with $K^0_{S}$ regeneration and so we revert to 
$K^0_{L},K^0_{S}$ representation. In the amplitude $M_d(K^0\rightarrow 
\pi \pi)$ the wave function of $K^0$ is given by (16). Since (15) and 
(16) are nearly equal, for (16) we use the relations identical 
to those for (15): 
\begin{eqnarray}
\left.\mid\!K^0\right>=(\left.\mid\!K^0_L\right>+\left.\mid\!K^0_S\right>)
/\sqrt{2},\nonumber\\
\left.\mid\!\bar{K}^0\right>=(\left.\mid\!K^0_L\right>-\left.\mid\!K^0_S\right>)
/\sqrt{2}.
\end{eqnarray}
Then 
\begin{eqnarray}
M_d(K^0\rightarrow \pi \pi)=[M_d(K^0_L\rightarrow 
\pi \pi)+M_d(K^0_S\rightarrow \pi \pi)]/\sqrt{2},\nonumber\\
M_d(\bar{K}^0\rightarrow \pi \pi)=[M_d(K^0_L\rightarrow 
\pi \pi)-M_d(K^0_S\rightarrow \pi \pi)]/\sqrt{2}.
\end{eqnarray}

Finally
\begin{equation}
M'(K^0_L\rightarrow \pi \pi )=M_d(K^0_L\rightarrow \pi \pi)+
\frac{\epsilon }{V}M_d(K^0_S\rightarrow \pi \pi).
\end{equation}
Here $(\epsilon /V)M_d(K^0_S\rightarrow \pi \pi)$ is the amplitude of 
regeneration followed by decay, $M_d(K^0_S\rightarrow \pi \pi)$ is 
the in-medium amplitude of the decay $K^0_S\rightarrow \pi \pi$. 
The factor $\epsilon /V$ is the same as in Eqs. (6) and (10) what is 
obvious since $K^0_L$ and $K^0_S$ are the superposition of $K^0$ and 
$\bar{K}^0$. 

The regeneration arises from difference between the $K^0N$- and 
$\bar{K}^0N$-intereactions which is included in $V$. Due to this we 
focused our attention on $V$- and $\Delta m$-dependences. The 
generalization for the relativistic case and the approach based on 
exact solution of equations of motion (see Sect. 5) will be presented 
in the following paper.

The calculation of process width is similar to that given by (11). We 
would like to emphasize the following. With the calculation of the 
total process probability, the result obtained by means of (23) should 
be multiplied by the factor
\begin{equation}
R=e^{-(\Gamma +\Gamma _a)t}
\end{equation}
which is conditioned by decay and absorption of $K^0$ ($\Gamma $ and 
$\Gamma _a$ are the widths of decay and absorption of $K^0$, respectively). 
These processes have been not included in $K^0_m(x)$ since the corresponding 
values are not used directly in the calculation of the $M'(K^0_L\rightarrow \pi \pi )$. 
In other words, the above-mentioned processes have been considered as 
background ones. The same is true for the process (1).

\section{Comparison with previous calculation}
The previous calculations (see, for example Eqs. (1) and (2) of Ref. [3] or 
Eq. (9.32) of Ref. [4] or Eqs. (7.83)-(7.89) of Ref. [5]) give 
\begin{eqnarray}
\left.\mid\!K'_L\right>_{st}=\left.\mid\!K_L\right>+r\left.\mid\!K_S\right>,\nonumber\\
r=i\pi N\Lambda f_{21}/[k(i\mu +1/2)]\sim f_{21}/(i\Delta m/\Gamma _S+1/2).
\end{eqnarray}
Here $\left.\mid\!K'_L\right>_{st}$ is the in-medium state expressed through the vacuum states. 
The notations are the same as in [5]. In the more recent papers (see, for 
example, Refs. [11,12]) we did not find an alternative calculations and 
relation different from (25).

To compare with our calculation we write (23) in the form
\begin{eqnarray}
M'(K^0_L\rightarrow \pi \pi )=<\pi \pi \mid H_W\mid\!K'_L>,\nonumber\\
\left.\mid\!K'_L\right>=\left.\mid\!K_L\right>+\frac{\epsilon }{V}\left.\mid\!K_S\right>,\nonumber\\
\frac{\epsilon }{V}=\frac{m}{4\pi N}\frac{\Delta m}{f_{21}}.
\end{eqnarray}
Here $\epsilon /V$ is the amplitude of regenerated $K^0_{S}$. Compared to (25),
the $f_{21}$- and $\Delta m$- dependences are inverse: $\epsilon /V\sim 1/r$.

The amplitude of any $ab$ transition is proportional to $\Delta m/V\sim \Delta m
/f_{21}$. The identical factor takes place for the $n\bar{n}$ transition in the 
medium followed by annihilation [8,9,13-15,17,18] and neutrino oscillations [16]. In 
(25) the inverse dependence takes place. It is important that $\Delta m$ and 
$f_{21}$ are the crucial values which define the process speed.

We also note that in the old calculation Eq. (21) is used as well (see Eq. (1) 
from Ref. [1]) since one cannot do without it. 
  
The fundamental difference between our and previous calculations lies in the 
process model. We write the starting Eqs. (3) from [2]:
\begin{eqnarray}
(\partial_x-ink)K^0=0,\nonumber\\
(\partial_x-in'k)\bar{K}^0=0,
\end{eqnarray}
where $n$ and $n'$ are the index of refraction for $K^0$ and $\bar{K}^0$, 
respectively. In notations of Ref. [2] $K^0=\alpha $ and $\bar{K}^0=\alpha '$,
$K^0_{S}=\alpha _1$ and $K^0_{L}=\alpha _2$. This follows from initial 
conditions [2]: $\alpha _2(0)=1$, $\alpha _1(0)=0$. 

There is no off-diagonal mass $\epsilon =(m_L-m_S)/2$. Equations (27) are non-coupled. 
The non-coupled equations exist only for the stationary states and don't exist for $K^0$ and $\bar{K}^0$.

In [2] the result (6) is obtained from (3) by means of change of  variables and introduction of $\tau $ ($\tau =1/\Gamma $). Indeed, in Eq. (27) of this paper we substitute $K^0=(\alpha _1+ i\alpha _2)/ \sqrt{2}$, $\bar{K}^0=(\alpha _1-i\alpha _2)/ \sqrt{2}$ and include the effect of the weak interactions as in [2]. We obtain Eq. (5) and result (6) from Ref. [2].

(Suppose that $\alpha $ and $\alpha '$ correspond to the stationary states 
$K^0_{S}$ and $K^0_{L}$. However, the index of refraction for $K^0_{L}$ 
and $K^0_{S}$ are unknown. Besides, in this case the values $\alpha _{1}(t)$ 
and $\alpha _{2}(t)$ obtained in [2] describe $K^0$ and $\bar{K}^0$ rather than  
$K^0_{S}$ and $K^0_{L}$.)

In [2] the off-diagonal mass term was omitted; $\Delta m$ appears due to the fact that $m_{\alpha _1}\ne m_{\alpha _2}$ in the corresponding plane waves. Equations (27) are due to the direct, erroneous analogy with ordinary optics (see paragraph before Eq. (1) of Ref. [2]). We deal with two non-stationary states (coupled equations).

Consequently, the term $H_{K^0\bar{K}^0}$ has been omitted. Unfortunately, from the Eq. (4) of the original paper [1] it is difficult to understand the simple fact given above. With the results of [1] in [2,3] the expression for $r$ (see Eq. (25) of this paper) was obtained. It was subsequently included in [4,5] and other papers. 

The reduction of the mass matrix to the diagonal form is only the change of 
variables. In the old calculations the term $H_{K^0\bar{K}^0}$ was omitted. This is 
fundamental error since it leads to a qualitative disagreement in the results. In our 
calculation the term $H_{K^0\bar{K}^0}$ involved in (8). In the approach given below (see 
Sect. 5) it is involved in the equations of motion.

\section{Approach based on exact solution of equations of motion}
Equation (23) was obtained in the first order in $\epsilon $. One can attempt to 
solve the problem exactly. By analogy with Eqs. (14) and (22) the in-medium wave 
functions of $K^0_{S}$ and $K^0_{L}$ are 
\begin{eqnarray}
K^0_L(t)=[K^0(t)+\bar{K}^0(t)]/\sqrt{2},\nonumber\\
K^0_S(t)=[K^0(t)-\bar{K}^0(t)]/\sqrt{2}.
\end{eqnarray} 
The exact solutions $K^0(t)$ and  $\bar{K}^0(t )$ of Eqs. (2) and (3) are 
substituted into (28). The value $\mid K^0_S(t)\mid ^2$ should be compared with 
the process probability corresponding to amplitude (23).

These calculations are rather cumbersome. There are non-trivial physical problems 
as well. For example, the optical potential was adapted for the Schrodinger-type 
equations. We deal with the system of equations (see [17] for more details). 
Equations (28) correspond to $K^0_S$ in the final state, whereas (23) includes the 
decay $K^0_S\rightarrow \pi \pi$ as well. 

We would like to emphasize the following. In the standard model of the $n\bar{n}$ 
transition based on equations of motion the optical theorem is used. This is wrong 
since the $S$-matrix is essentially non-unitary [17,18]. (For the oscillations in the 
external field [19-22] the Hamiltonian is Hermitian and so there is no similar 
problem.) In the calculation given above the optical theorem is not used. 

\section{Conclusion}
In the old model [1-5] the $K^0(t)$ and $\bar{K}^0(t)$ are described by (27), which 
is wrong. Equations (27) are due to the erroneous analogy with ordinary optics [2]. 
$K^0$ and $\bar{K}^0$ correspond to the non-stationary states and coupled equations 
of motion. Consequently, in the previous calculations the problem different from 
regeneration has been solved. 

In this paper and Ref. [23] we propose the alternative model which is typical for the theory of the multistep processes.
It is significant that width of $K^0_{L}K^0_{S}$ transition (17) obtained in [23] coincides with Eq. (28) from [24], which 
can be considered as a test. Our results differ fundamentally from previous one.

The calculation by means of perturbation theory is useful on the following reason. The exact calculation given in [23] is cumbersome and formal. The calculation given above has a transparent physical sense. It is also significant 
that in the approach based on diagram technique one can avoid the potential description of 
$\bar{K}^0$-medium interaction (see [8], for example). This is essential since the potential 
description is crude due to the strong absorption of $\bar{K}^0$.

In the approach based on exact solution of equations of motion [23] we start from 
(2) and (3) and not (27). In Sect. 3 of this paper we make the same in the framework of 
perturbation theory. The Hamiltonian and initial state wave function are given by 
(8) and (16), respectively. The perturbation theory in $\epsilon $ is used. The 
parameter $\epsilon $ is extremely small. The regeneration amplitude involves  
all the orders in $V$ and $H_W$ and the first order in $H_{K^0\bar{K}^0}$.

We do not overestimate the role of regeneration. $\Delta m$ is extracted from 
free-space oscillations without recourse to $V$, i.e., potentials of $K^0$ and 
$\bar{K}^0$. Nevertheless, in any case the process should be described correctly.
We will continue our consideration in the following paper.

\end{document}